# Sol-gel synthesis and multiferroic properties of pyrochlore-free Pb(Fe$_{0.5}$Nb$_{0.5}$)O$_3$ thin films

L. Imhoff [1], M. B. Di Marco[1], S. Barolin[1], M. A. Rengifo[3,4], M.H. Aguirre[2,3,4] and M.G. Stachiotti[1]

*1-Instituto de Física Rosario (IFIR), CONICET – Universidad Nacional de Rosario. 27 de Febrero 210 Bis, Rosario, Argentina.*
*2-Dept. de Física de la Materia Condensada, Universidad de Zaragoza, Pedro Cerbuna, 12, 50009, Zaragoza, Spain.*
*3-INMA- Instituto de Nanociencia y Materiales de Aragón, CSIC-Universidad de Zaragoza, Mariano Esquillor s/n, 50018, Zaragoza, Spain.*
*4-Laboratorio de Microscopías Avanzadas, Universidad de Zaragoza, Mariano Esquillor s/n, 50018, Zaragoza, Spain.*

**Abstract**

Lead iron niobate (PbFe$_{0.5}$Nb$_{0.5}$O$_3$ - PFN) thin films were synthesized by a modified sol-gel route, which offers the advantage of a rapid, simple and non-toxic reaction method. Polycrystalline perovskite-structured PFN thin films without pyrochlore phases were obtained on Pt/Ti/SiO$_2$/Si substrates after sintering by rapid thermal annealing at 650 °C. TEM and AFM images confirmed the excellent quality of the sintered film, while EDS spectroscopy revealed the presence of oxygen vacancies near the film/substrate interface. Electric measurements show good dielectric properties and ferroelectric behavior, characterized by typical C-V curves and well-defined P-E ferroelectric loops at 1 kHz, with remanent polarization values of ~12 µC/cm$^2$. The polarization, however, increases with decreasing frequency, indicating the presence of leakage currents. I-V measurements show a significant increase in DC-conduction at relatively low fields (around 100 kV/cm). The films display ferromagnetic behavior at room temperature, with magnetic remanence around 30 emu/cm$^3$ and a coercive field of 1 kOe. These values are significantly higher than those obtained for PFN powders fabricated by the same sol-gel route, as well as the magnetization values reported in the literature for epitaxial films.

## Introduction

Lead iron niobate, PbFe$_{0.5}$Nb$_{0.5}$O$_3$ (PFN), is an AB´B´´O$_3$ perovskite-type material and is well known for being one of the first studied multiferroics. It was originally reported by Smolenskii et al. in the late 1950s [1]. Since then, and especially in the beginning of this century, many works were done about the fabrication of this material using different techniques. Multiferroicity implies the coexistence of at least two ferroic orders. In the case of PFN, the presence of magnetic Fe$^{3+}$ cations at the B-site of the perovskite structure makes possible the existence of magnetic ordering through the super-exchange interaction between iron and oxygen ions Fe$^{3+}$-O$^{2-}$-Fe$^{3+}$. On the other hand, ferroelectric properties arise from the strong hybridization between B (Fe and Nb) d and O p orbitals [2], and from the stereochemically active lone pair electrons of Pb$^{2+}$, relevant for ferroelectricity in Pb-based perovskites [3]. The early works by Bokov et al. [4] indicate that PFN presents a paraelectric to ferroelectric phase transition at 385 K, and becomes antiferromagnetically ordered at 145 K. These transition temperatures allow this material to become a multiferroic below 145 K. In spite of that, there are multiple studies reporting ferromagnetic behavior at room temperature, both in bulk [5,6] and epitaxial thin film [7,8] form. For that reason, PFN is still a topic of extensive research, not only due to the interest in its fundamental understanding, but also due to the possibility of finding a magnetoelectric coupling at room-temperature. Such a discovery would open chances to a huge variety of new applications in the field of memory and logic devices.

There is controversy around the crystalline structure of PFN samples. The room-temperature phase has been reported to be either monoclinic *Cm* [5,6] or rhombohedral *R3m* [9,10]. A recent work on PFN ceramics by Bartek et al. [11] proposes a coexistence of mixed monoclinic and rhombohedral phases, and it concludes that the monoclinic phase is dominant, and the rhombohedral phase is only favored at higher sintering temperatures. On the other hand, Carpenter et al. [12] modeled the x-ray pattern of PFN ceramics using a cubic perovskite structure, and measurements using x-ray absorption spectroscopy (XAS) showed that the shell structure around Fe atoms exhibits tetragonal local symmetry at room temperature [13].

Most of the studies reported on the PFN system have focused on ceramic and single crystal samples, while investigations of thin film are scarcer, despite the potential for a wide variety of applications in micro and nanodevices. The few works reporting multiferroic behavior in PFN thin films involve epitaxial samples fabricated by pulsed laser deposition (PLD) [8,14–16]. There have also been some studies on thin films based on chemical sol-gel techniques [17–20], but they do not report magnetic studies. For example, Sedlar et al. [17] fabricated sol-gel films using rapid thermal annealing as a sintering method, obtaining samples with good dielectric and ferroelectric properties. However, the authors did not report any details of the solution synthesis nor magnetic characterization of the samples. Furthermore, there was no mention of the phase characterization of the synthesized films. It is worth mentioning that the formation of a pure perovskite phase in PFN presents a real challenge due to the stabilization of pyrochlore phases ($A_2B_2O_7$, $A_2B_2O_6$, $Pb_3Nb_4O_{13}$) at low temperatures, which has been a significant drawback in multiple sample processing methods [5,11,14]. This difficulty is particularly important in sol-gel thin film processing, where the synthesis of pure perovskite films on platinized silicon substrates has remained a difficult task due to the tendency towards the formation of pyrochlores. However, pyrochlore-free PFN films were obtained by using other substrates, such as silica glass [14] or $SrTiO_3$ [17]. Platinized substrates, on the other hand, offer easy integrability into semiconductor devices, and research towards the fabrication of single-phase PFN films by chemical routes on these kind of substrates holds great importance.

Traditional sol-gel routes make use of toxic 2-methoxyethanol as a solvent and involve refluxing mechanisms during the preparation of precursor solutions. Such practices have been common in the previous works done on PFN thin films by sol-gel [18–20]. Another issue frequently found in the fabrication of lead-based oxides is the high volatility of Pb during sintering, especially for thin films, due to their high evaporation surface. Lead deficiency can modify the system's stoichiometry and is among the main factors that promote the formation of pyrochlore phases [21]. The main strategy for overcoming this problem has been the addition of an excess of Pb in the precursor solutions. However, if the Pb excess is too high, it makes the Pb-compound dissolution process difficult and also favors the formation of other Pb-rich secondary phases. Su et al. [22], for example, fabricated Pb(Fe,Ta)$O_3$ films using different amounts of lead up to a 100% excess. They found that the films prepared with high lead excess displayed amorphous and crystallized lead oxide phases. Kang et al. [18] obtained pyrochlore-free PFN films on silica glass using a 38% lead-excess. Liu et al. [20] fabricated pure perovskite films onto SrTiO$_3$ substrates with 20% lead-excess and a PbO-rich environment. They found that, in order to obtain a pure perovskite phase, a PbO-rich environment must be used during post-heat-treatment because the excess of Pb in the starting composition is not enough to overcome the effects of lead oxide evaporation, as well as the pyrochlore phase formation during sintering. The films were then heat-treated in a range from 650 ºC to 720 ºC for 10 h in a PbO-rich atmosphere, which is not suitable for practical applications.

In this work, we report the sol-gel synthesis and characterization of pyrochlore-free PFN thin films deposited on platinized silicon substrates. We employed a modified sol-gel route that avoids the use of 2-methoxyethanol through the addition of a chelating agent that reduces the reactivity of starter alkoxides. This one-pot synthesis method is very simple and does not require the use of distillation or refluxing strategies. We show that PFN films thermally treated at 650º C by rapid thermal annealing exhibit a pure perovskite phase, without any pyrochlore secondary phases. We investigate and discuss the microstructural, electric and magnetic properties of the films.

**Experimental Procedure**

The sol-gel route used in this work is based on a previously developed route for the synthesis of (1-x)PZT-xPFN (x≤0.5) thin films [23]. We employed acetoin (3-hydroxy-2-butanone, $CH_3COCH(OH)CH_3$, Aldrich) as a chelating agent, along with the following reactive precursors for metallic cations: lead (II) acetate trihydrate ($Pb(CH_3CO_2)_2·3H_2O$, 99+% Aldrich), iron acetylacetonate ($Fe(C_5H_7O_2)_3$ 97% Aldrich) and Niobium ethoxide ($Nb(OCH_2CH_3)_5$ 99.5% Aldrich). The details of the solution preparation process were already published and can be found in reference [23]. Nevertheless, the route is summarized in the flow chart shown in Fig. 1.

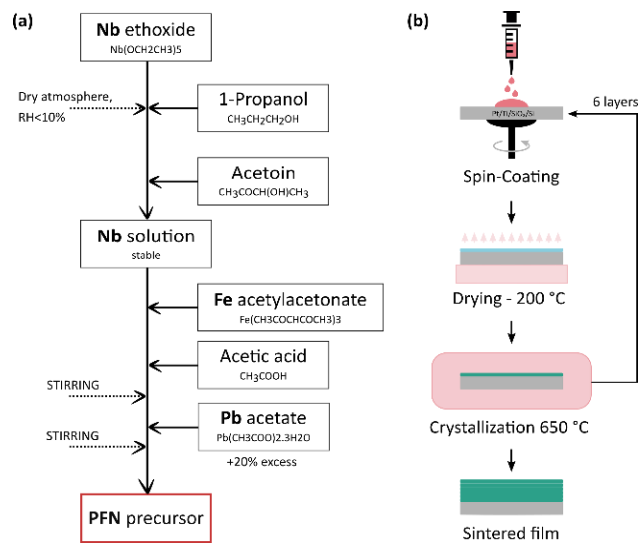

**Fig. 1. (a)** Flow chart for the PFN solution synthesis route and **(b)** scheme of the deposition process and sintering of thin film.

The concentration used for spin-coating deposition was 0.2 M. In contrast to the case shown in reference [23], we found it necessary to apply a 1:4 first diluted layer to achieve a good adhesion between the substrate and the PFN solution. The precursor solution was deposited at 4000 rpm for 15" onto Pt/Ti/SiO$_2$/Si wafers, followed by drying at 200 °C for 5' on a heating plate. The samples were thermally treated by rapid thermal annealing (RTA): each layer was quickly heated-up to 650 °C (heating rate 50 °C/s), kept for 1 min at that temperature, and then quenched in air. The process was repeated 6 times to obtain 6-layered films, and no final heating was performed. All these steps were found to be optimal for the fabrication of (1-

x)PZT-xPFN thin films [23], since RTA conditions were crucial to avoid the pyrochlore phases formation and obtain a full perovskite microstructure.

An equivalent powder was obtained by drying the solvent from the PFN precursor solution. Drying was performed at 90 °C in a muffle (air atmosphere) for a minimum of 12 hours. Part of the dried gel-powder was left for DTA/TGA analysis, and most of it was calcinated at 850 °C for 4 hours to obtain a single perovskite phase. Because of the excess mass in the powder compared to the film, a higher temperature and a longer calcination time were required in this case to avoid the formation of secondary phases, such as lead, iron and niobium oxides, as well as pyrochlore phases.

Simultaneous Differential Thermal Analysis (DTA) and Thermogravimetric Analysis (TGA) measurements were carried out on the as-obtained gel powder from room temperature to 800 °C in a normal atmosphere using a Shimatzu DTG 60 H equipment with a heating rate of 10 °C/min. The sintered samples were characterized by X-ray diffraction using a Phillips X'pert Pro X-Ray diffractometer with Cu Kα radiation of wavelength 1.5406 Å. The measurements for the films were taken in the grazing incident configuration with an incident beam angle of 5°, 2θ varying between 20° and 60° and a scan rate of 0.02 °/s. The powders were measured in Bragg-Brentano configuration, with 2θ varying between 20° and 90°, 0.01° step size. Raman spectra were acquired with a Renishaw inVia Reflex micro-Raman spectrometer by means of the 532nm Ar-ion laser line (~16mW nominal power). The surface morphology of the films was observed by Atomic Force Microscopy (Nanotec ELECTRONIC AFM equipment) in tapping mode using a silicon probe with 150 kHz resonance frequency and a constant force of 7.4 N/m. Focused Ion Beam (FIB) was used to prepare a cross-section sample of the multi-layered film by FEI equipment Dual Beam Helios Nanolab 650. The cross-section specimen was then analyzed by Transmission Electron Microscopy (TEM) at 300 kV using a FEI Tecnai F30 and a FEI Titan3 coupled to a HAADF-STEM detector with probe-aberration correction. Chemical composition was analyzed by Energy Dispersive X-ray Spectroscopy (EDS). For electrical measurements, 0.25 mm diameter platinum top electrodes were deposited by DC-sputtering on the surface of the films. Dielectric properties were measured using an Agilent 4294A LCR meter. The ferroelectric loops were obtained at room temperature from a Sawyer-Tower circuit applying alternate signals at 50 Hz and 1 kHz frequencies. I-V curves were obtained using a Keithley 2635A Source Measure Unit. The magnetic measurements were carried out in a commercial superconducting quantum interference device magnetometer (SQUID).

**Results and discussion**

A transparent dark red homogeneous PFN precursor solution was obtained, which remained stable for weeks after preparation, being stored at 6 °C. Density and viscosity measured values for the 0.2 M PFN solution at 25 °C were (0.91 ± 0.01) g/cm$^3$ and (2.06 ± 0.01) cP, respectively, and these values did not present considerable changes within 2 months. The solutions were stable in air after being prepared. The DTA/TGA curves in Fig. 2 show the transformation process of the amorphous gel powder to a fully crystallized powder. The TGA curve displays the weight loss due to decomposition and burning of the residual organic groups, which is around 35 % at 500 °C.

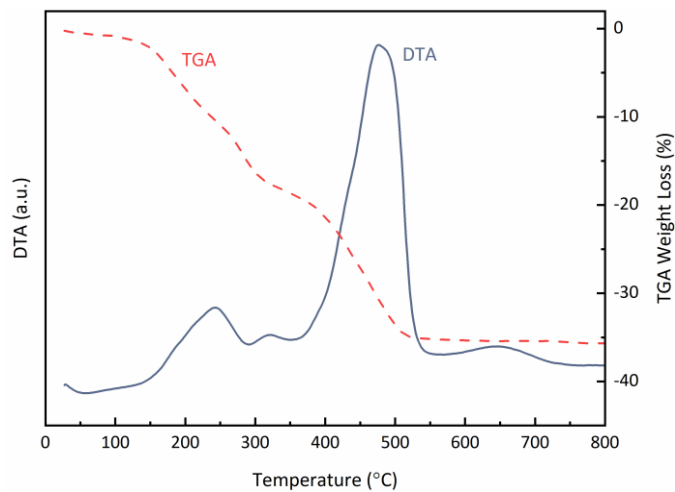

**Fig. 2.** DTA/TGA analysis of the dried powder obtained from the PFN precursor solution.

No weight loss is observed above that temperature. The DTA curve displays an intense exothermic peak centered near the temperature where the elimination of the residual organic groups finishes (480 °C). This peak is attributed to the elimination of the last organic residues with a simultaneous onset of the crystallization.

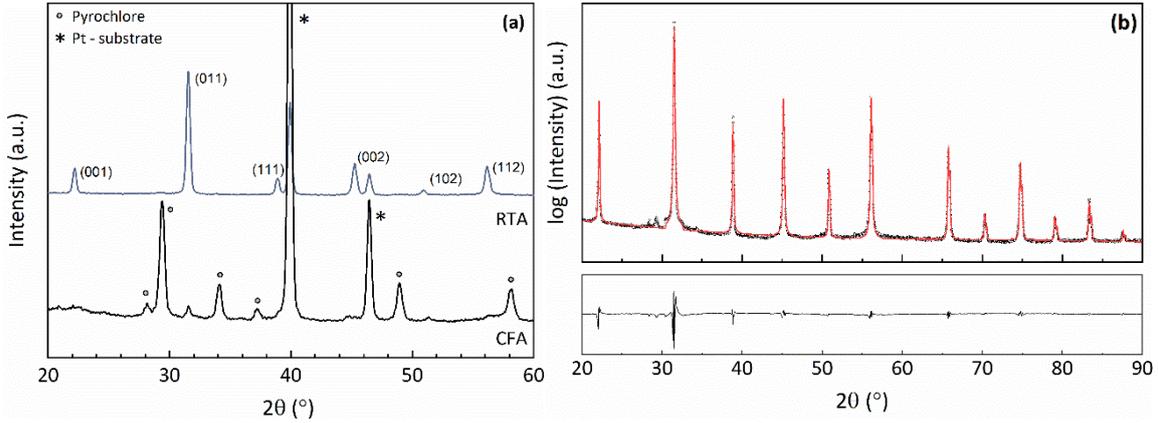

**Fig. 3. (a)** The gray XRD curve shows perovskite indexed peaks for the RTA film. The black one corresponds to the film treated by CFA and presents the pyrochlore peaks (marked with a circle). **(b)** Rietveld refinement of the XRD perovskite pattern of the powder sample. The difference between experimental and refined data is shown at the bottom.

The XRD pattern of the PFN film treated at 650 °C by RTA is shown in Fig. 3a**.** Perovskite peaks were indexed using a cubic structure as a reference. For comparison, we also show the pattern of a film treated by conventional furnace annealing (CFA) at the same temperature (each layer was introduced for 3 minutes in a preheated furnace at 650 °C, and the multi-layer coating was finally annealed for 15 min and then quenched in air). It is clear that RTA treatment produces a polycrystalline single-phase perovskite PFN thin film, without pyrochlore phases. Conversely, the sample treated by CFA presents a small amount of perovskite and mostly pyrochlore phases. Rietveld refinement was performed for the RTA sample using the software MAUD [24]. Three different fittings, enforcing tetragonal, monoclinic, or rhombohedral symmetries, were performed, and the corresponding structural data is presented in Table 1. The refinement quality parameters indicate a good fit between the three proposed structures and the experimental data. Among them, a slightly better fit is achieved for the monoclinic phase. We also analyzed the XRD pattern of the equivalent powder calcined at 850 °C, presented in Fig.

3b. The pattern revealed the formation of a single-phase randomly-oriented perovskite structure. We evaluated the three different symmetries and once again we found a better fit for the monoclinic phase, in agreement with the results reported in previous works [6,11]. Finally, the fitting parameters for the three tested structures are all very good and similar, indicating that it is difficult to distinguish which pseudocubic symmetry fits better, within the detection limits of the employed XRD equipment.

| Sample | Phase | Lattice Parameters | | | Angles | | Tetragonality | Quality parameters | |
|---|---|---|---|---|---|---|---|---|---|
| | | a (Å) | b (Å) | c (Å) | α (°) | β (°) | c/a | Sigma | $R_{rw}$ (%) |
| Film | Monoclinic (Cm) | 5.667 (5) | 5.664 (6) | 4.028 (2) | -- | 89.73 (3) | -- | 1.14 | 12.4 |
| | Rhombohedral (R3m) | 4.006 (1) | -- | -- | 90.15 (2) | -- | | 1.17 | 12.8 |
| | Tetragonal (P4mm) | 3.997 (1) | -- | 4.012 (2) | -- | -- | 1.004 | 1.21 | 13.2 |
| Powder | Monoclinic (Cm) | 5.6747 (1) | 5.6725 (1) | 4.0169 (1) | -- | 89.90 (1) | -- | 5.6 | 5.8 |
| | Rhombohedral (R3m) | 4.0131 (4) | -- | -- | 90.00 (1) | -- | -- | 6.1 | 6.3 |
| | Tetragonal (P4mm) | 4.0127 (2) | -- | 4.0147 (2) | -- | -- | 1.0005 | 6.2 | 6.0 |

**Table 1.** All calculated lattice parameters for the PFN film and powder, and quality fitting factors.

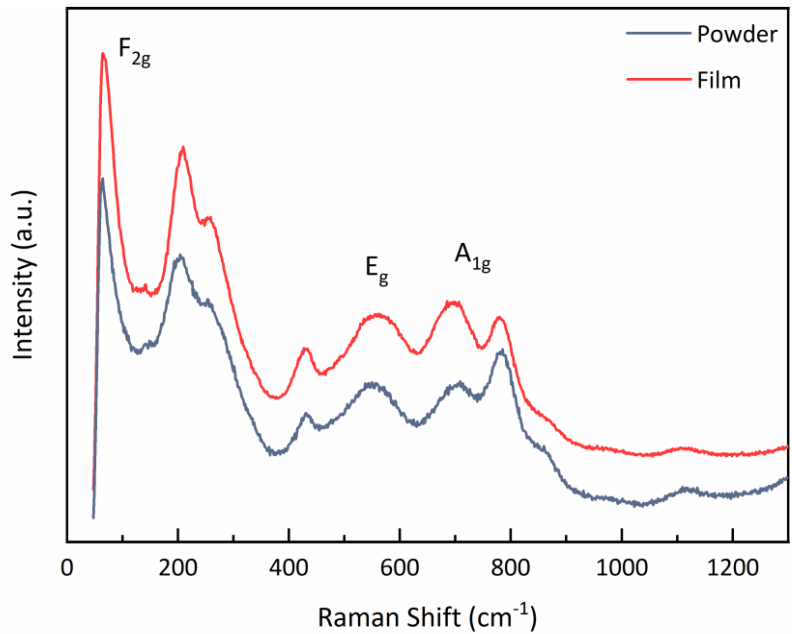

**Fig. 4.** Room-temperature RAMAN spectra for PFN thin film (red) and powder (gray).

Raman measurements were made for both thin film and powder samples and the room-temperature spectra are shown in Fig. 4. The spectrum of the film contains all the bands present in the spectrum of the powder, both displaying well-defined sharp phonon peaks. The frequencies of the majority of the bands correspond well to the values reported in previous

studies [25–27], and are similar in the film (64, 208, 256, 429, 560, 694, 780, 860 and 1100 cm$^{-1}$) and in the powder (64, 203, 256, 429, 550, 702, 782, 860 and 1100 cm$^{-1}$) samples. Note that some frequencies in the film are slightly shifted as compared to the powder, probably due to strain effects caused by the substrate.

The *Fm-3m* symmetry is considered as a good approximation for describing the Raman spectra of PFN and other pseudocubic ferroelectrics [25–27]. This symmetry corresponds to the ideal periodic alternation of two types of cations at the B-site position (Fe and Nb) in a cubic perovskite lattice. For this symmetry, the group theory predicts four Raman active modes $A_{1g}+E_g+2F_{2g}$, and the $4F_{1u}$ infrared (IR)-active modes. However, additional peaks become visible in Raman spectra due to local structural distortions with respect to the ideal lattice. For instance, ionic disorder and off-center displacements lead to a local symmetry breaking, and it is well established that Raman spectra of disordered relaxor materials contain more lines than expected for the ideal symmetry of the crystal structure. According to this, and comparing with the mentioned reports done on this material, we can assume that the lower frequency band (around 64 cm$^{-1}$) is attributable to $F_{2g}$ mode, which corresponds to Pb displacements. The bands at 208 cm$^{-1}$ and 256 cm$^{-1}$ probably arise from local symmetry breaking, like $BO_6$ octahedron tilting and/or polar distortions owing to cations off the center shifts. The Raman active $E_g$ mode corresponds to the experimental peak at 560 cm$^{-1}$. The two high-frequency Raman bands at 694 cm$^{-1}$ and 780 cm$^{-1}$ originate from the $A_{1g}$ mode of the $BO_6$ octahedra and correspond to Fe-O and Nb-O stretching vibrations. The shoulder observed at 860 cm$^{-1}$ is also relative to this B-O bond, while the band at 1100 cm$^{-1}$ is likely to be a second-order feature.

An AFM image of the PFN thin film surface is shown in Fig. 5, displaying a surface area of 1500x1500 nm. The grain size presents a positively-skewed normal distribution with a mode value of 60 nm (as shown in the inset graphics). The larger values probably correspond to the agglomeration of smaller grains, which cannot be resolved by means of the measurement conditions. Rugosity (RMS) was calculated from a 2x2 µm image giving a value of ~31 nm.

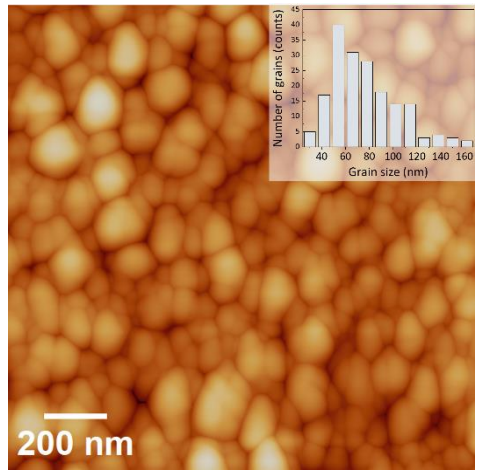

**Fig. 5.** AFM image of the surface morphology for the PFN thin film and grain size distribution histogram (inset).

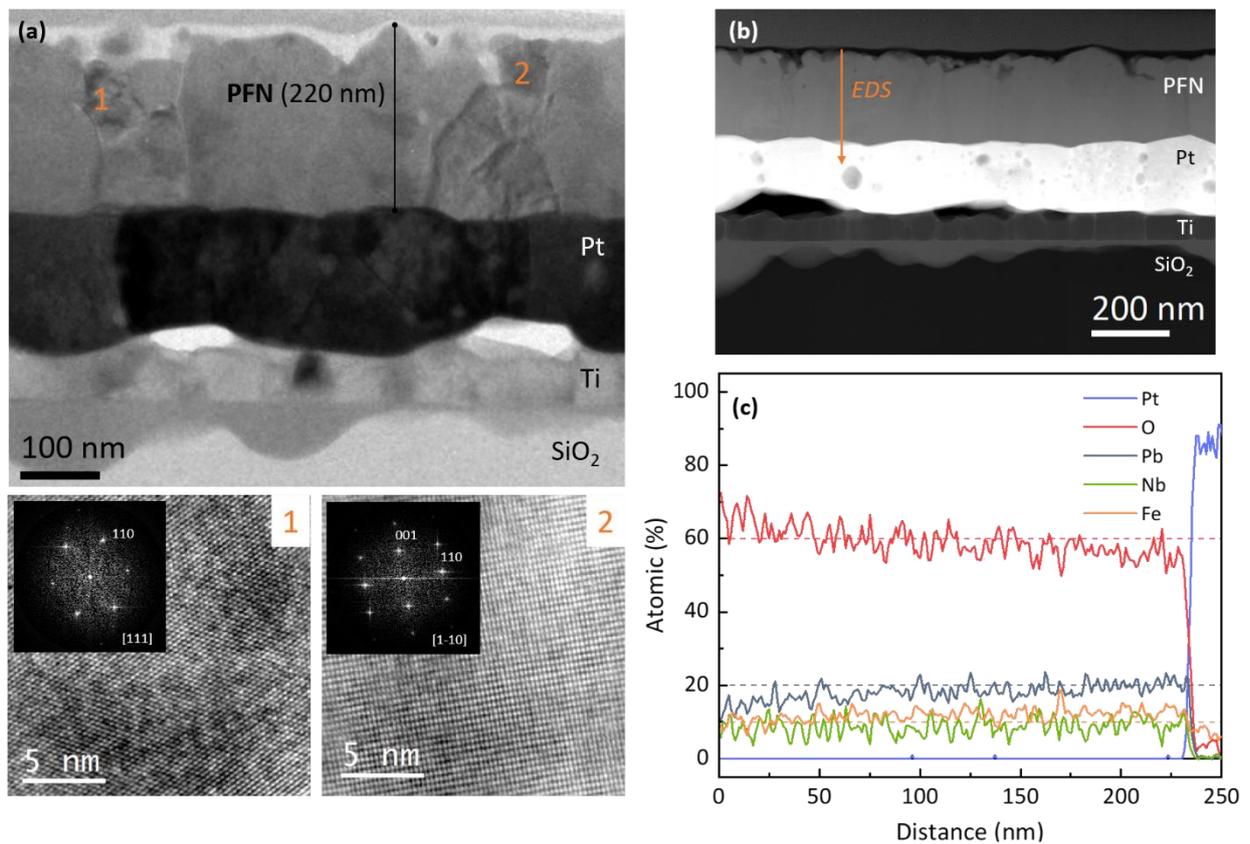

**Fig. 6. (a)** Cross-sectional TEM image of the thin film sample, showing the substrate layers and the PFN film. Also, two different grain orientations with the corresponding FFTs, indexed according to the pseudocubic perovskite structure. In figures **(b)** and **(c)** a STEM-mode profile image and the EDS line spectra are shown.

In Fig. 6a, a bright field TEM image taken in a cross-sectional sample of PFN is shown. The layers corresponding to the film and the substrate are properly labeled. The film thickness was measured and found to be (220±10) nm. High-resolution images of two grains and their corresponding FFT, revealing two different crystalline orientations, are also presented in Fig. 6a. In Fig. 6b a STEM-mode image of the profile is shown. A significant rugosity is observed on the sample, which correlates well with the high RMS rugosity found by AFM measurements.

The EDS-profile displayed in Fig. 6c, obtained by measuring along the sample thickness, shows a good agreement between the concentration of the elements present in the sample and the corresponding stoichiometry for PFN. However, it is noticeable that the region near the bottom electrode presents an oxygen deficiency of around 5%, indicating the presence of vacancies. A lead deficiency is also observed near the surface of the sample, as expected due to evaporation.

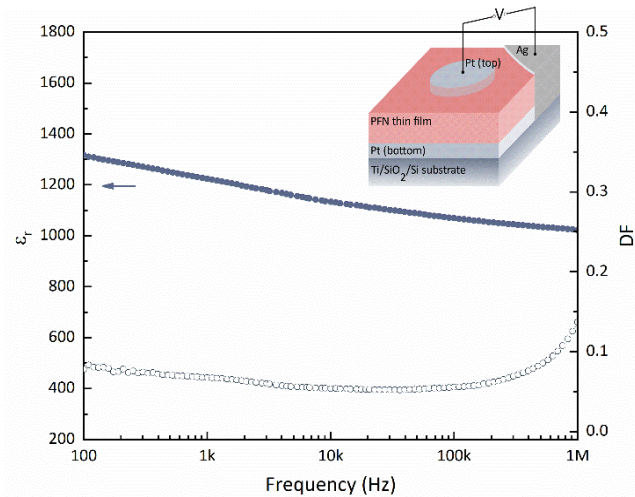

**Fig. 7.** Dielectric permittivity and loss factor as a function of frequency under an applied AC voltage. Schematic diagram of the Pt/PFN/Pt measured devices. Pt was used as a bottom and top electrode (inset).

The complete set of electric measurements was conducted using a bottom-top configuration, wherein positive voltages were applied to the top platinum contact, and negative voltages were applied to the bottom platinum contact. A schematic representation of the studied devices can be seen in the inset of Figure 7.

The dielectric properties of the film were first investigated in terms of the dielectric constant ($\varepsilon_r$) and dissipation factor (DF), which were measured as a function of frequency in the range between 100 Hz and 1 MHz. The results are shown in Fig. 7. The figure shows that both the dielectric constant and the dielectric loss steadily decrease with increasing frequency. For instance, $\varepsilon_r$ decreases from ~1300 at 100 Hz to ~1000 at 1 MHz. Such frequency dispersion is a characteristic property of relaxor ferroelectrics. We note that relaxor ferroelectric characteristics were observed in epitaxial thin films [8,15]. The dissipation factor, on the other hand, displays a flatter behavior, with DF values around 0.05 even at low frequencies, indicating the absence of space-charge polarization contributions at low applied voltages.

The ferroelectric characterization was performed by three different measurement methods on the micro-devices shown in figure 7 (inset). Capacitance–Voltage (C-V) characteristics were measured at three different frequencies and the results are plotted in Fig. 8a. The bias voltage was increased from 0V to positive values, then switched to negative values, and then back to zero. Ferroelectric behavior is evidenced by the splitting of peaks in these curves, which is directly related to the irreversible switching of ferroelectric polarization [28]. The C-V curves show an imprint towards positive voltages for all measurement frequencies, as evidenced by the non-coincidence of the capacitance values at 0V (DC bias). Ferroelectric features were also analyzed from polarization-electric field (P-E) hysteresis loops measured at different frequencies using a Sawyer-Tower circuit (Fig. 8b). The loop at 1 kHz is well-shaped, though it is not completely saturated at high fields. At lower frequencies, the ferroelectric behavior is still evidenced but the leakage increases, causing the loop to open. The remanent polarization and coercive field values measured at 1 kHz are ~12 $\mu C/cm^2$ and 25 kV/cm, respectively. It is worth noting that a leaky behavior has also been observed in other reports of P-E ferroelectric loops for PFN samples, whether in bulk [6] or thin films [16,17], and a good saturation is typically not achieved. In sol-gel synthesized films, for instance, a saturation in polarization values was never fully reached up to voltages close to electrical breakdown [17]. The highest obtained values of remanent polarization and coercive field for those films were 4.5 µC/cm² and 40 kV/cm, respectively. Gao et. al. reported $P_r$ = 7.4 µC/cm² and $E_c$ = 10.5 kV/cm measured in films fabricated by PLD [16].

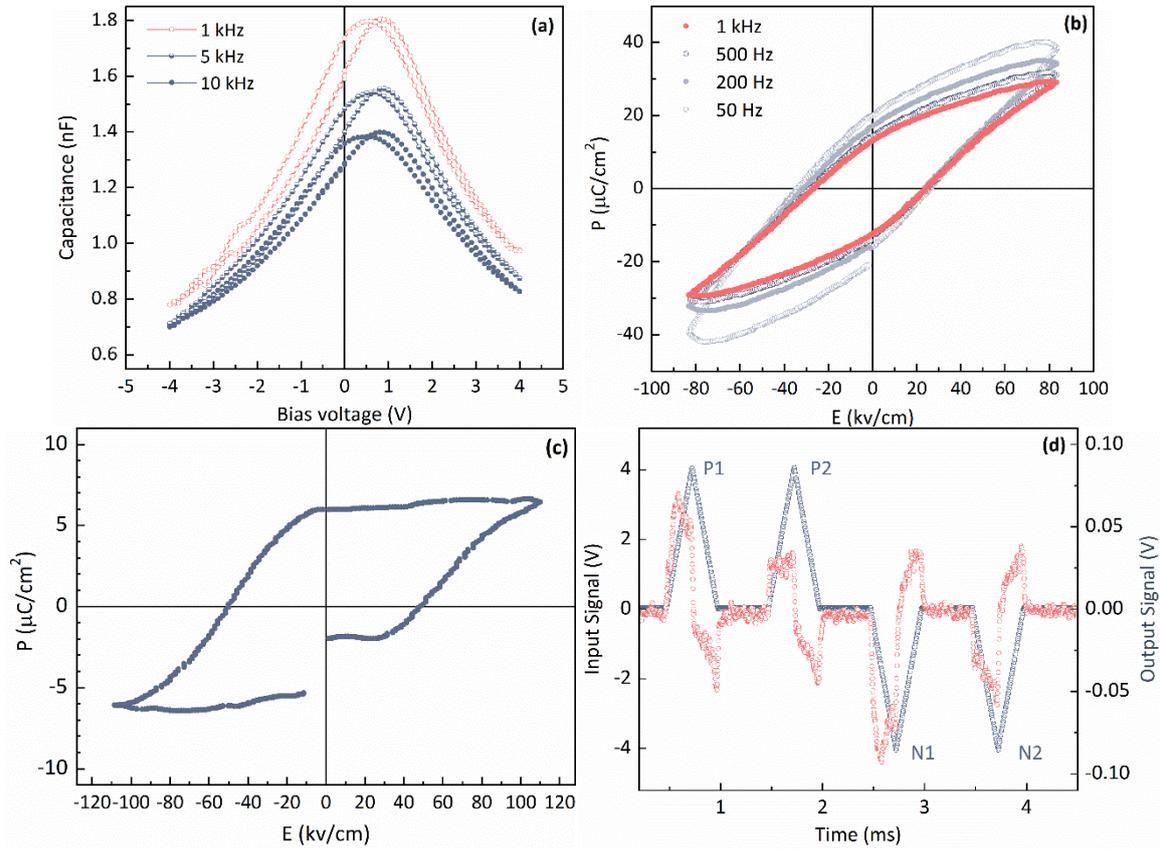

**Fig. 8. (a)** Capacitance-Voltage curves measured at 3 different frequencies. **(b)** P-E loops obtained for different frequencies from a conventional Sawyer-Tower circuit. **(c)-(d)** PUND loop measured at 1kHz and the corresponding applied signals.

It is also noteworthy that in our samples, achieving better saturation is not possible, as higher fields cannot be applied without causing dielectric rupture. We estimate for the sol-gel PFN film a breakdown field of ~100 kV/cm, much smaller than the one observed for the solid solution between PFN and PZT [23]. Low breakdown fields were also reported in other works done in bulk samples [6,16] and films [16].

PUND (Positive-Up-Negative-Down) pulses are an appropriate technique to distinguish ferroelectric polarization from conductive contributions and obtaining a more reliable remanence value. We tested our samples by applying consecutive unipolar pulses with an electric voltage of 3 and 4 V with 1 kHz frequency (corresponding time of 1 ms between pulses, see Fig. 8d). The first P1 and N1 pulses are used to fully align the FE domains. During the next two P2 and N2 pulses, the conductive and dielectric contributions arise. By removing these, only FE behavior remains, represented by the squared loops shown in Fig. 8c. In this measurement, an important contribution of leakage is observed (Fig. 8d). The remanent polarization obtained by this technique has a value of ~7 $\mu C/cm^2$.

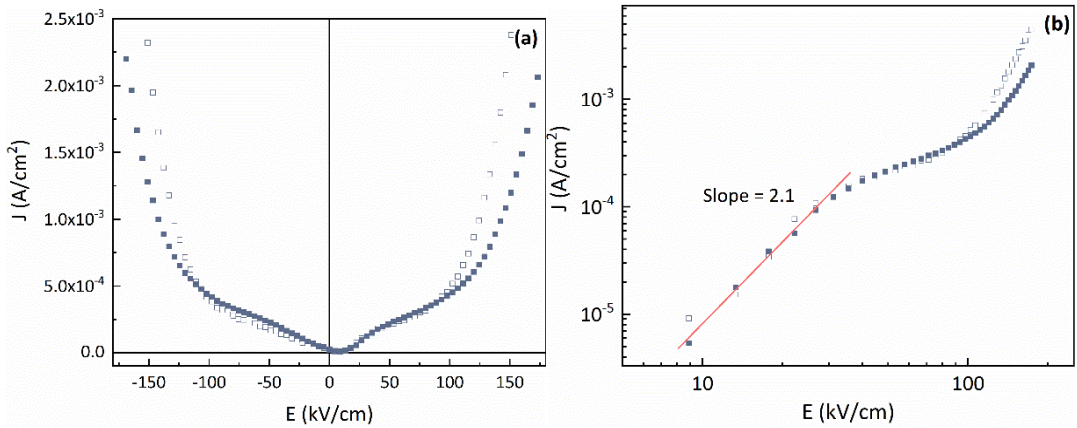

**Fig. 9. (a)** Current-Voltage characteristics as J vs E for a ferroelectric capacitor before (empty square) and after (filled square) the P-E hysteresis measurement. **(b)** log (J) vs log (E) plot.

Current-voltage (I-V) curves were obtained by applying consecutive voltage pulses with a width of 10 ms, and a delay time of 250 ms between -4 and 4 V. Fig. 9a shows the current density as a function of the applied field for a ferroelectric capacitor before and after the P-E measurement. The curves are not symmetric, and a small imprint towards positive fields, similar to that present in the C-V curves, is observed. The current shows an increase for fields up to 30 kV/cm and a sharp increase after an inflection point around 100 kV/cm. It is noticeable that this point coincides with the maximum field that can be applied to measure P-E hysteresis loops at different frequencies without causing dielectric breakdown. Sedlar et al. in reference [17] reported a similar current increase at high fields. Furthermore, they found an ohmic type of conduction, as evidenced by the linear behavior of the current observed for low electric fields (slope=1). When plotting log (J) vs log (E), as shown in Fig. 9b, we also found a linear behavior at low fields, but with a slope value around 2. This behavior can be associated with the presence of space-charge-limited currents, represented by a modified Child's law [29]. On the other hand, a linear behavior at low fields is also observed when plotting ln(J) versus $E^{1/2}$, which is compatible with a Schottky emission mechanism [30]. We note that the space-charge-limited current conduction mechanism is supported by the EDS-profile presented in Fig. 6d, which shows the presence of oxygen vacancies distributed over a region near the PFN-substrate interface and lead vacancies near the surface of the film [29]. Nevertheless, it is difficult to determine which is the dominant conduction mechanism in the PFN sample, also considering the polycrystalline nature of the film and the wide distribution of grain sizes. Additional measurements should be necessary to further clarify this point, i.e., measuring leakage curves with varying delay times and temperature.

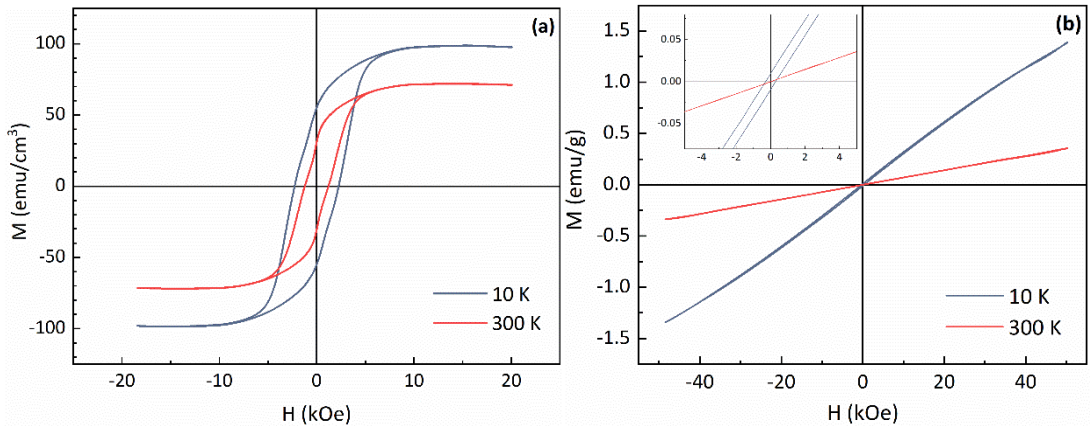

**Fig. 10.** M-H loops measured at 10 K (gray) and 300 K (red) in the PFN **(a)** thin film and **(b)** equivalent powder.

The dependence of the in-plane magnetization values with the magnetic field (M-H loops) was obtained at two different temperatures and shown in Fig. 10a. In both cases, a clear ferromagnetic behavior is observed, with weak remanent magnetization ($M_r$) values around 55 and 31 emu/cm$^3$ for 10 K and 300 K, respectively. The coercive field ($H_c$) values were measured to be 2300 and 1200 Oe. In contrast, magnetization values reported for PLD-deposited epitaxial thin films are typically lower than 10 emu/cm$^3$ (for instance, M$_r$ ~5 emu/cm$^3$ and H$_c$ ~200-300 Oe [7,8]). For comparison, the magnetic behavior of the equivalent powder was also investigated (Fig. 10b). In this case, the remanence at room-temperature is zero, and a very small opening of the loop is observed at low temperatures (see the inset graphic). This is in agreement with the results found in the literature for powder and pellet samples [5,31].

The magnetic behavior of PFN is still a subject of intense debate [32,33]. Some authors propose that room-temperature weak ferromagnetism may arise from clustering and spin canting of the Fe ions [5,8,34]. This is a possible explanation, because the crystal structure of PFN is monoclinic and Fe$^{+3}$–O–Fe$^{+3}$ spins would be no longer collinear. As a result, a small canting of the spins yields weak ferromagnetic behavior. On the other hand, Peng et. al. [7] proposed the presence of super-antiferromagnetic clusters as a plausible scenario to explain weak ferromagnetism observed up to 300 K. In our case, the presence of oxygen vacancies near the PFN-substrate interface (Fig. 5d) could explain the strong increase of the magnetization remanence observed in the sol-gel film, as compared to the equivalent powder and other results reported in the literature for PFN ceramics [5,6] and epitaxial films [7,8]. We suggest the existence of a ferromagnetic exchange mechanism, which involves spin-polarized electrons trapped at oxygen vacancies. It is expected that oxygen vacancies create Fe$^{3+}$– V$_O$ – Fe$^{3+}$ groups in the crystal structure of PFN, where V$_O$ denotes the oxygen vacancy. Since the electronic configuration of the Fe$^{3+}$ cation only has unoccupied minority spin orbitals (3d$^5$), the trapped electron will be polarized down and the two neighbor Fe$^{3+}$ ions up. This direct ferromagnetic coupling, called F-center exchange (FCE) mechanism [35], was used to explain room-temperature ferromagnetism in magnetic-doped oxides [36–38]. We thus conclude that the FCE coupling could be responsible for the magnetization increased values in the sol-gel film.

**Conclusions**

Polycrystalline PFN thin films with a thickness of 220 nm were fabricated using a simple, non-toxic and rapid chemical solution synthesis method. The films, deposited on platinized silicon substrates, present single-phase perovskite structure, with no detectable pyrochlore phases. An equivalent pyrochlore-free powder was obtained from the dried chemical precursor solution for a parallel study. Raman spectroscopy confirmed that both the film and powder samples exhibit pseudocubic symmetry, and Rietveld's analysis showed that the monoclinic *Cm* phase gives the best fitting of the XRD data. TEM and AFM cross-sectional and topographic images confirmed the excellent quality of the sintered film. EDS spectroscopy revealed a noticeable presence of oxygen vacancies distributed over a region near the film/substrate interface and the I-V behavior of the film was associated with a space-charge-limited current conduction mechanism. In spite of that, good dielectric and ferroelectric properties with well-defined hysteresis loops were obtained at different frequencies, with remanent polarization values around 12 $\mu C/cm^2$ at 1 kHz. Magnetic characterization at room temperature showed weak ferromagnetic behavior in the film, with magnetization values significantly higher than those previously reported for epitaxial PFN films. These results offer a novel approach for synthesizing good-quality polycrystalline PFN thin films, contributing to the fundamental understanding of the material, and opening new possibilities for applications in multifunctional micro-devices.

**Acknowledgements**

We thank Dr. Jamal Belhadi (LPMC, Université de Picardie) for the Raman measurements.

**Funding**

This work was supported by *Consejo Nacional de Investigaciones Científicas y Técnicas de la República Argentina* (CONICET) by PIP N° 0374. M.G.S. thanks support from *Consejo de Investigaciones de la Universidad Nacional de Rosario* (CIUNR). We acknowledge the financial support of the European Commission by the H2020-MSCA RISE projects MELON (Grant Nº 872631) and ULTIMATE-I (Grant Nº 101007825).

**Data and code availability**

Data will be made available on request.

**Declaration of generative AI in scientific writing**

During the preparation of this work, the authors used ChatGPT (Chat Generative Pre-Trained Transformer) in order to enhance the manuscript's readability and writing quality. After using this tool, the authors reviewed and edited the content as needed and take full responsibility for the content of the publication.

**Declaration of competing interest**

The authors declare that they have no known competing financial interests or personal relationships that could have appeared to influence the work reported in this paper.

**Author Contributions**

**L. Imhoff:** writing, data curation, electric and magnetic measurements. **M. B. Di Marco:** sample preparation, review & editing. **S. A. Barolin:** XRD acquisition and analysis, AFM images, review & editing. **M. A. Rengifo:** software, review & editing. **M. H. Aguirre:** TEM imaging and analysis, review & editing. **M. G. Stachiotti:** writing, conceptualization, supervision.

**Captions**

**Fig. 1. (a)** Flow chart for the PFN solution synthesis route and **(b)** scheme of the deposition process and sintering of thin film.

**Fig. 2.** DTA/TGA analysis of the dried powder obtained from the PFN precursor solution.

**Fig. 3. (a)** The gray XRD curve shows perovskite indexed peaks for the RTA film. The black one corresponds to the film treated by CFA and presents the pyrochlore peaks (marked with a circle). **(b)** Rietveld refinement of the XRD perovskite pattern of the powder sample. The difference between experimental and refined data is shown at the bottom.

**Table 1.** All calculated lattice parameters for the PFN film and powder, and quality fitting factors.

**Fig. 4.** Room-temperature RAMAN spectra for PFN thin film (red) and powder (gray).

**Fig. 5.** AFM image of the surface morphology for the PFN thin film and grain size distribution histogram (inset).

**Fig. 6. (a)** Cross-sectional TEM image of the thin film sample, showing the substrate layers and the PFN film. Also, two different grain orientations with the corresponding FFTs, indexed according to the pseudocubic perovskite structure. In figures **(b)** and **(c)** a STEM-mode profile image and the EDS line spectra are shown.

**Fig. 7.** Dielectric permittivity and loss factor as a function of frequency under an applied AC voltage. Schematic diagram of the Pt/PFN/Pt measured devices. Pt was used as a bottom and top electrode (inset).

**Fig. 8. (a)** Capacitance-Voltage curves measured at 3 different frequencies. **(b)** P-E loops obtained for different frequencies from a conventional Sawyer-Tower circuit. **(c)-(d)** PUND loop measured at 1kHz and the corresponding applied signals.

**Fig. 9. (a)** Current-Voltage characteristics as J vs E for a ferroelectric capacitor before (empty square) and after (filled square) the P-E hysteresis measurement. **(b)** log (J) vs log (E) plot.

**Fig. 10.** M-H loops measured at 10 K (gray) and 300 K (red) in the PFN **(a)** thin film and **(b)** equivalent powder.